\documentstyle[12pt]{article}

\title{
\begin{flushright}
{\large Yaroslavl State University \\
        Preprint YARU-HE-97/04 \\
        hep-ph/9708293} \\[10mm]
\end{flushright}
       {\LARGE\bf About an influence of external electromagnetic fields} \\
       {\LARGE\bf on the axion decays}}

\author{{\Large\bf N.V.~Mikheev } \\[2mm]
        {\large\it
             Yaroslavl State University, Yaroslavl 150000, Russia} \\
        {\large\it E-mail: mikheev@yars.free.net} \\[4mm]
        {\Large\bf and} \\[4mm]
        {\Large\bf L.A.~Vassilevskaya} \\[2mm]
        {\large\it
             Moscow Lomonosov University, V-952, Moscow 117234, Russia} \\
        {\large\it E-mail: vasilevs@vitep5.itep.ru}}

\date{}

\begin{document}

\maketitle

\begin{abstract}
Some results of an external electromagnetic field influence
on axion decays are presented. It is shown that the field
catalyzes substantially these processes. The result describing
the axion decay $a \to f \tilde f$ we have obtained can be of 
use  to re-examine the lower limit on the axion mass in case 
of possible existence of strong magnetic fields at 
the early Universe stage.
\end{abstract}

\thispagestyle{empty}

\newpage

Various extensions of the Standard Model predict the existence
of Nambu-Goldstone bosons, which are the subject of constant 
theoretical and experimental investigations. 
The existence of such particles would be a
low-energy manifestation of new physics at energy scales much larger
than can be probed in the laboratory and 
they could play an important role in astrophysics and cosmology.

In this work we present some of our results concerning the
decays of the most widely discussed pseudo-Goldstone boson, the
axion~\cite{P1,WW}, associated with Peccei-Quinn $PQ$-symmetry
(see, for example~\cite{P2,P3}).

The allowed range for the axion mass, and therefore for
the $PQ$-symmetry breaking scale, $f_a$, is strongly constrained 
by astrophysical and cosmological considerations~\cite{P3,Raf1,Tur}
which leave a rather narrow window 
\footnote{The current status of the axion window is reviewed in
the recent paper by Raffelt~\cite{Raf2}.}:

\begin{equation}
10^{-6} {\rm eV} \leq m_a \leq 10^{-3} {\rm eV}.
\label{eq:MA}
\end{equation}

\noindent The upper bound on $m_a$ arises from astrophysical 
arguments because axion emission removes energy from stars 
altering their evolution. If the axion has a mass near 
the lower limit, then axions play a very important role in the 
Universe and could be the Universe's cold dark matter.
A survey of various processes involving weakly coupled particles
production and astrophysical methods for obtaining constraints 
on the parameters of axion models is given in
a wonderful book by Raffelt~\cite{Raf3}.

Despite constant theoretical and experimental 
interest to the axion all attempts of its experimental 
detection have not a success yet. It can be naturally
explained by the fact that axions are very light, very weakly 
interacting (coupling constant $\sim f^{-1}_a$) and very long living.
The axion lifetime in vacuum is gigantic one:

\begin{equation}
\tau (a \to 2\gamma) \sim 6,3 \cdot 10^{48} \,s \,
\left ( {10^{-3} eV \over m_a } \right )^6 \; 
\left ( {E_a \over 1 MeV } \right ) .
\label{eq:T0}
\end{equation}

On the other hand, investigations in strong external 
electromagnetic fields become of principal interest in astrophysics 
and cosmology of the early Universe in that they can be effective 
in getting new limits on weakly coupled particles parameters. 
At present by strong magnetic fields we mean the fields of strength 
of order and more the critical, so called, Schwinger value
$B > B_e$, $B_e = m^{2}_{e}/e \simeq 0.44 \cdot 10^{14} \, G$.
It should be noted that
by now astrophysical knowledge of strong magnetic fields
which can be realized in the nature have changed essentially.
At present the field strengths inside the astrophysical objects
as high as $10^{15} \div 10^{17} G$~\cite{B-K} are discussed.

It is known that an external field can catalyze substantially 
the decay processes. For example, in our recent work~\cite{L1}
we have studied a double radiative axion decay 
$a \to \gamma \gamma$ in the external field. 
The effect of significant enhancement of the decay probability 
by the field was discovered, namely, the field removes 
the main suppression caused by the smallness of the axion mass. 
To illustrate this we present here the comparison of the decay
probability we have obtained with the one in vacuum:

\begin{eqnarray}
R_e = {W^{(F)} \over W^{(0)}}  \simeq 10^{37} 
\left ( {\cos^2\beta \over E/N - 1.92} \right )^2 \;
\left ( {10^{-3} eV \over m_a } \right )^4 \; P(\chi).
\nonumber
\end{eqnarray}

\noindent 
Here $\cos^2 \beta$ determines the electron Yukawa coupling 
$g_{ae} = \frac {1}{3}\cos^2\beta (m_e/f_a)$ in the DFSZ 
model~\cite{DFSZ};
$P(\chi)$ is the function of the field dynamical parameter. 
Notice that this result (the strong catalyzing influence 
of the field) wasn't curious or unexpected one.
In our previous works~\cite{L2} we have studied the other loop
process, namely, the neutrino radiative decay in external electromagnetic
field of various configurations and found the analogous phenomenon.
Unfortunately despite such an enhancement the axion lifetime in this
decay the field remains the big one.

The other special feature of the external field influence
is that the field can open new channels forbidden in vacuum 
(for example, the well known photon splitting into 
electron-positron pair $\gamma \to e^+ e^-$~\cite{Klep} 
and into two photons $\gamma \to \gamma \gamma$~\cite{Adl}). 

A tree level axion decay into charged fermion-antifermion pair 
$a \to f \tilde f$ is one of the processes which is forbidden
in vacuum when $m_a < 2 m_f$ and opened in the magnetic field.
This is due to the fact that the kinematics of a charged 
particle in a magnetic field is that which allows to have 
both time-like and space-like total momentum for 
the charged fermion-antifermion pair.
For this reason this process becomes possible in the magnetic
field even for massless pseudoscalar particle as it occurs in 
photon splitting $\gamma \to e^+ e^-$.

An amplitude of this decay can be immediately 
obtained from the Lagrangian of axion-fermion interaction:

\begin{equation}
{\cal L}_{af} =  \frac{ g_{af}}{2 m_f} 
(\bar f \gamma_\mu \gamma_5 f) \, \partial_\mu \, a,
\label{eq:L}
\end{equation}

\noindent using the known solutions of the Dirac equation in a 
magnetic field. Here $g_{af} = C_f m_f/f_a$ is a dimensionless 
coupling constant, $C_f$ is a model-dependent factor,
$m_f$ is a fermion mass. To obtain the decay probability one 
has to carry out a non-trivial integration over the phase 
space of the fermion pair taking the specific kinematics of 
charged particles in the magnetic field into account.

However the decay probability can be obtained by another easier way. 
It is known, that the decay probability of  
$a \to f \tilde f$ is connected with the imaginary part of the 
amplitude of the $a \to f \tilde f \to a$ transition via the 
fermion loop by the unitarity relation:

\begin{equation}
E_a W(a \to f \tilde f) = Im \,{\cal M}_{a \to a}.
\label{eq:W0}
\end{equation}

\noindent 
In the second order of the perturbation theory the 
field-induced part of ${\cal M}_{a \to a}$ 
can be presented in the form:

\begin{eqnarray}
\Delta {\cal M}_{a \to a} & \simeq &  \frac{g^2_{af}}{8 \pi^2} 
\cdot \beta \, q^2_\perp \,
\int\limits_0^1 d u \, \int\limits_0^\infty d t \, 
e^{-i \Phi(u,t)} \,
{\cos \beta t - \cos \beta ut \over \sin \beta t},
\label{eq:M}\\
\Phi(u,t) & = & 
\left (
m_f^2 - q^2_\perp \,{1 - u^2 \over 4} 
\right ) t 
+ {q^2_\perp \over 2 \beta} \,
{\cos \beta ut - \cos \beta t \over \sin \beta t},
\nonumber\\
q^2_\perp & = & \frac{e_f^2}{\beta^2} (q F F q) 
\simeq E_a^2 \sin^2 \theta,
\nonumber
\end{eqnarray}

\noindent where $\beta = \vert e_f B \vert = \sqrt {e_f^2(FF)/2}$;
$e_f = e Q_f$, $e > 0$  is the elementary charge,
$Q_f$ is a relative fermion charge; $F_{\alpha \beta}$ is the
external magnetic field tensor
($(q F F q) = q_\mu F^{\mu \nu} F_{\nu \rho} q^\rho $, 
$( F F ) =  F_{\mu \nu} F^{\nu \mu} $); 
$E_a$ is the energy of the decaying axion;
$\theta$ is the angle between the vectors 
of the magnetic field strength ${\vec B}$ and the momentum of
the axion ${\vec q}$. 
\noindent We stress that in order to obtain the correct expression
for ${\cal M}_{a \to a}$ one has to use a derivative nature of the 
axion-fermion interaction Lagrangian~(\ref{eq:L}) as it was 
first emphasized by Raffelt and Seckel~\cite{Raf4}.

Here we present the results of our calculations in a 
more intersting limiting case, $ |e_f B | > q^2_\perp$, when 
the charged fermion-antifermion pair can be born only in the states 
corresponding to the lowest Landau level. This becomes possible 
if the high energy axion propagates almost along the magnetic 
field strength vector $\sin^2\theta \sim 4 m^2_f/ E^2_a$, 
so $E^2_a \sin^2\theta \sim 4 m^2_f < | e_f B | $.
The decay probability has the form:

\begin{equation}
W \simeq \frac{g^2_{af}}{4 \pi} \, 
\frac{\beta}{E_a} \,
\frac{
\exp \left ( {\mbox{\large $
-\frac{ q^2_\perp}{2 \beta}$}}
\right )}
{\left({\mbox{\large $ 1 - 
\frac{4 m^2_f}{q^2_\perp }$}} \right)^{1/2}} \; \rho
\label{eq:W1}
\end{equation}

\noindent Here the factor $\rho$ is introduced which takes 
into account a possible influence of the medium on the axion decay 
($\rho = 1$ when the medium is absent):

\begin{eqnarray}
\rho  =  \frac {\int d W \, ( 1 - n_f ) \, ( 1 - n_{\tilde f} )}
{\int d W} ,
\label{eq:Rho}
\end{eqnarray}

\noindent where 

\begin{eqnarray}
n_f = \left [ exp \left ( \frac{E_f - \mu}{T} \right ) + 1
\right ]^{-1}
\nonumber
\end{eqnarray}

\noindent is the Fermi-Dirac distribution function of fermion at
a temperature~$T$, $\mu$ is the chemical potential,
$n_{\tilde f}$ is the distribution function of antifermions.

\noindent 
The expression for the lifetime is:

\begin{equation}
\tau \simeq \frac{1,4 \cdot 10^7}{\rho} \, 
\frac{1}{\vert Q_f \vert} \,
\left ( {10^{-13} \over g_{af}} \right )^2 \;
\left ( {10^{15} G\over B} \right ) \;
\left ( {E_a \over 100 MeV } \right ) \; 
\left( 1 - \frac{4 m^2_f}{q^2_\perp} \right)^{1/2} \; s.
\label{eq:T1}
\end{equation}

\noindent As it follows from Eq.~(\ref{eq:T1}), the axion 
lifetime tends to zero under the kinematical condition:
$q^2_\perp \to 4 m^2_f$. Actually the expression
$\left( 1 - 4 m^2_f /q^2_\perp  \right)^{1/2}$
has a minimal nonzero value due to the axion dispersion
in the magnetic field:

\begin{equation}
\left(
1 - \frac{4 m^2_f}{q^2_\perp}
\right)^{1/2}_{min}
\simeq \frac{1}{\sqrt{3}} \,
\left ( \frac{g^2_{af}}{2 \pi} \frac{B}{B_f} \right )^{1/3}.
\label{eq:SIN}
\end{equation}

\noindent So, the minimal value for the axion lifetime is:

\begin{equation}
(\tau)_{min} \simeq \frac{1,7 \cdot 10^{-2}}{\rho} \, 
\left ( {1 MeV \over \vert Q_f \vert m_f} \right )^{2/3} \;
\left ( {10^{-13} \over g_{af}} \right )^{4/3} \;
\left ( {10^{15} G\over B} \right )^{2/3} \;
\left ( {E_a \over 100 MeV } \right ) \; s .
\label{eq:TM}
\end{equation}

Considering possible applications of the result we have 
obtained to cosmology it is necessary to take an influence 
of a hot plasma into account. It is known that such an influence is
reduced to suppressing statistical factors in the integral 
over the phase space of the fermion pair.
Under the early Universe conditions the hot plasma is 
nondegenerated one ($\mu \ll T$) and the medium parameter 
$\rho$ in~(\ref{eq:W1}),~(\ref{eq:T1}) and~(\ref{eq:TM})
is close to $\frac{1}{4}$. 

The result we have obtained could be of use to re-examine
the lower limit of the axion mass in the case of a possible 
existence of strong magnetic fields ($B > B_e$) at the early 
Universe stage.

\end{document}